\documentclass[psamsfonts,12pt]{article}

\usepackage[mathscr]{eucal}
\usepackage{graphicx}
\usepackage{amsthm}
\usepackage{amsmath}
\usepackage{latexsym}
\usepackage{amscd}
\usepackage{amsfonts}
\usepackage{amssymb}
\usepackage{graphicx}
\usepackage{graphicx}
\date{}

 \usepackage{mathptmx}      
 \usepackage{latexsym}
 \usepackage{booktabs}
 \usepackage{multirow}
 \usepackage{tabularx}
 \usepackage{caption}
 \usepackage[figuresright]{rotating}
 
 \usepackage{graphicx}
 \usepackage[mathscr]{eucal}
 \usepackage{amsmath,amssymb,latexsym,amscd}
 \usepackage{multirow}
 
 \usepackage{mathptmx} 
 \usepackage{latexsym}
 \usepackage{booktabs}
 \usepackage{multirow}
 \usepackage{tabularx}
 \usepackage{caption}
 \usepackage[figuresright]{rotating}

\RequirePackage{fix-cm}
\usepackage{graphicx}
\usepackage{float}

\usepackage{amssymb}
\usepackage[mathscr]{eucal}

\def\C{\mathbb C}

\mathchardef\ssneq="3B28

\def\m@th{\mathsurround=\z@}

\begin{document}

\title{ Applying Adomian Decomposition Method to Solve Burgess Equation with a Non-linear Source  }
\author{O. Gonz\'alez-Gaxiola\footnote{ogonzalez@correo.cua.uam.mx},  R. Bernal-Jaquez\\ ￼Departamento de Matem\'aticas Aplicadas y Sistema,s\\ Universidad Aut\'onoma Metropolitana-Cuajimalpa,\\C.P. 05300 Mexico, D.F., Mexico} \maketitle


\begin{abstract}
 
\noindent In the present work we  consider the mathematical model that describes brain tumour growth (glioblastomas) under medical treatment.  Based  on the medical study presented by  R. Stupp et al. (New Engl Journal of Med 352: 987-996, 2005)  which evidence that, combined therapies such as, radiotherapy and chemotherapy, produces negative tumour-growth, and using the mathematical model of P. K. Burgess et al. (J Neuropath and Exp Neur 56: 704-713, 1997) as an starting point, we present a model for tumour growth under medical treatment represented by a non-linear partial differential equation that is solved using the Adomian Decomposition Method (ADM). It is also shown that the non-linear term efficiently models the effects of the combined therapies. By  means of a  proper use of  parameters,  this model could be used for calculating doses in radiotherapy and chemotherapy.

\end{abstract}



\section*{Introduction}
\label{s:Intro}
\begin{figure} [h]
	\begin {center}
	\includegraphics[width=0.4\textwidth]{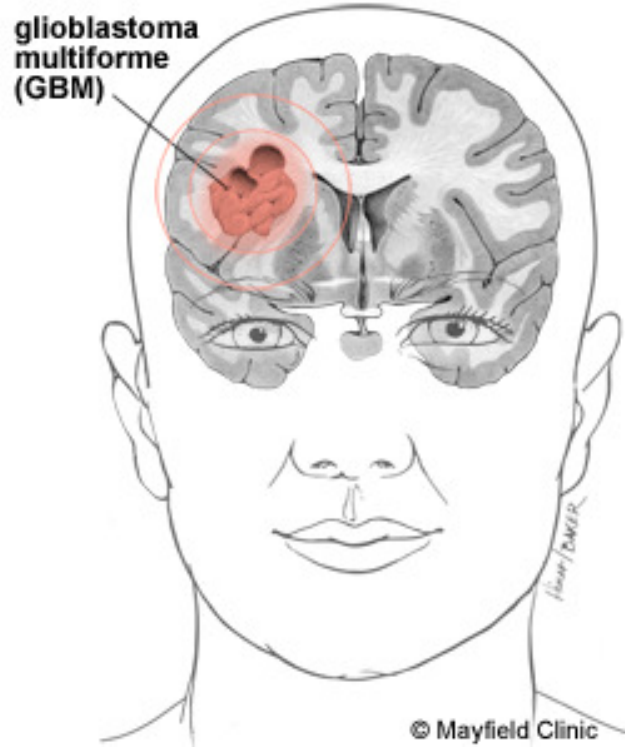}
	\caption{\small  Illustration of a glioblastoma tumour in the parietal lobe. \label{F1} }
	\end {center}
\end{figure}

\noindent The glioblastoma, also known as  glioblastoma multiforme (GBM), is a highly invasive glioma in the brain \cite{Sch}. It is the most common and most aggressive brain tumour in humans. From a medical point of view, GBM is a fast growing tumour made up of an heterogeneous mixture of poorly differentiated astrocytes, with pleomorphism, necrosis, vascular proliferation and high rate mitosis. This glioma can appear at any age but is more frequent among adults older than 45 years. Usually, they appear in the cerebral hemispheres but they could also appear in the cerebellum. From a mathematical point of view, they can be considered to have a spherical geometry as it is illustrated in figure \ref{F1} see \cite{May}. In 1997 P. K. Burgess  {\it et. al.}  proposed a 3-dimensional mathematical model that describes the growth of a glioblastoma free of any medical treatment  that could grow with no restrictions. This model provides information about the density change  of the tumour in any spatiotemporal point  but does not  give any information about the case in which some annihilation of  tumour cells could appear due, possibly, to the administration of cancericidal  substances and hence does not study the   dynamics of proliferation-annihilation of gliomas. It is worthy to say that some bi-dimensional mathematical  models preceded the  Burgess model as the ones formulated in \cite{Trac} y \cite{Wood}.

\noindent In the present work, and taking  the Burgess model as an starting point,  we will formulate a mathematical model that takes into account the action of some cancericidal substances  (as temozolomide and chemotherapy)  and hence the possibility to annihilate or diminish the growth  of the gliomas.
Our resulting model, in agreement with clinical data \cite{Roger},  is expressed in terms of a partial non-linear differential equation that is solved using the Adomian Decomposition Method  \cite{ADM1}, \cite{ADM2}.
The  model proposed  also allows  to compare the profile of  a tumour growing without any treatment with the  profile of a tumour subject to treatment, i,e., our model includes a term  that gives the difference  between the growth and annihilation of the glioma. Calibration of doses using this model as a basis  could result in the lengthening of life for glioma patients  \cite{Roger}.

\section*{Analysis of the Method}
\label{sn:AM}
\noindent  Adomian Decomposition Method (ADM) is a technique  to solve ordinary and partial nonlinear differential equations. Using this method, it is possible to express analytic solutions in terms of a rapidly converging series  \cite{ADM1}. In a nutshell,  the method identifies and separates the linear and nonlinear parts of a differential equation. By inverting and applying the highest order differential operator that is contained in the linear part of the equation, it   is possible to express the solution  in terms of the the rest of the equation affected by this inverse operator.  At this point, we propose to express this solution by means of  a decomposition series 
with terms that will be well determined by recursion and that gives rise to the solution components.  The nonlinear part is expressed in terms of the Adomian polynomials. 
The initial or the boundary condition and the terms that contain the independent variables will be considered as the initial  approximation. In this way and by means of a  recurrence relations, it is possible to calculate the  terms of the series by recursion that gives the approximate solution of the differential equation.\\

\noindent Given a partial (or ordinary) differential equation 

\begin{equation}
	Fu(x,t)=g(x,t)\label{eq:y1}
\end{equation}
with the initial condition 
\begin{equation}
	u(x,0)=f(x),\label{eq:y2}
\end{equation}
where $F$ is  differential operator  that could itself, in general, be nonlinear and therefore includes linear and nonlinear terms.\\  
In general, equation (\ref{eq:y1}) is be  written as 
\begin{equation}
	L_{t}u(x,t)+Ru(x,t)+Nu(x,t)=g(x,t)\label{eq:y3}
\end{equation}
where $L_{t}=\frac{\partial}{\partial t}$, $R$ is the linear remainder operator that could include partial derivatives with respect to $x$,   $N$ is a nonlinear operator which is presumed to be analytic and  $g$ is a non-homogeneous term that is independent of the solution $u$.\\
Solving for $L_{t}u(x,t)$, we have 
\begin{equation}
	L_{t}u(x,t)= g(x,t)-Ru(x,t)-Nu(x,t)\label{eq:y4}.
\end{equation}
As $L$ is presumed to be invertible, we can apply $L_{t}^{-1}(\cdot)=\int_{0}^{t}(\cdot)dt$ to both sides of equation  (\ref{eq:y4}) obtaining
\begin{equation}
	L_{t}^{-1}L_{t}u(x,t)= L_{t}^{-1}g(x,t)-L_{t}^{-1}Ru(x,t)-L_{t}^{-1}Nu(x,t).\label{eq:y5}
\end{equation}
An equivalent expression to  (\ref{eq:y5}) is
\begin{equation}
	u(x,t)= f(x)+L_{t}^{-1}g(x,t)-L_{t}^{-1}Ru(x,t)-L_{t}^{-1}Nu(x,t),\label{eq:y6}
\end{equation}
where $f(x)$ is the constant of integration with respect to $t$ that satisfies  $L_{t}f=0$. In equations where the initial value  $t = t_0$,  we can conveniently define  $L^{-1}$.\\
The ADM  proposes a decomposition series solution   $u(x,t)$ given as
\begin{equation}
	u(x,t)= \sum_{n=0}^{\infty}u_{n}(x,t).\label{eq:y7}
\end{equation}
The nonlinear term $Nu(x,t)$ is given as
\begin{equation}
	Nu(x,t)= \sum_{n=0}^{\infty}A_{n}(u_{0},u_{1},\ldots, u_{n})\label{eq:y8}
\end{equation}
where   $\{A_{n}\}_{n=0}^{\infty}$ is the Adomian polynomials sequence  given by (see deduction in appendix at the end of this paper)
\begin{equation}
	A_n= \frac{1}{n!}\frac{d^{n}}{d\lambda^{n}}[N(\sum_{k=0}^{n}\lambda^{k}u_{k})]\rvert_{\lambda=0}.\label{eq:y9}
\end{equation}
Substituting  (\ref{eq:y7}), (\ref{eq:y8}) y (\ref{eq:y9}) into equation (\ref{eq:y6}), we obtain
\begin{equation}
	\sum_{n=0}^{\infty}u_{n}(x,t)= f(x)+L_{t}^{-1}g(x,t)-L_{t}^{-1}R\sum_{n=0}^{\infty}u_{n}(x,t)-L_{t}^{-1}\sum_{n=0}^{\infty}A_{n}(u_{0},u_{1},\ldots, u_{n}),\label{eq:y10}
\end{equation}
\noindent with $u_0$ identified as $f(x)+L_{t}^{-1}g(x,t)$, and therefore, we can write
\begin{eqnarray*}
	u_{0}(x,t) &=& f(x)+L_{t}^{-1}g(x,t),\\
	u_{1}(x,t) &=& -L_{t}^{-1}Ru_{0}(x,t)-L_{t}^{-1}A_{0}(u_{0}),\\
	&\vdots \\
	u_{n+1}(x,t) &=& -L_{t}^{-1}Ru_{n}(x,t)-L_{t}^{-1}A_{n}(u_{0},\ldots,u_{n}).
\end{eqnarray*}
From which we can establish the following recurrence relation, that is obtained in a explicit way for instance in reference  \cite{Wen},
\begin{equation}
	\left\{
	\begin{array}{ll}
		u_{0}(x,t)=f(x)+L_{t}^{-1}g(x,t),\\
		u_{n+1}(x,t)=-L_{t}^{-1}Ru_{n}(x,t)-L_{t}^{-1}A_{n}(u_{0},u_{1},\ldots, u_{n}),\; n=0,1,2,\ldots.
	\end{array}
	\right.\label{eq:y11}
\end{equation}
Using  (\ref{eq:y11}), we can obtain an approximate solution of (\ref{eq:y1}), (\ref{eq:y2}) as  
\begin{equation}
	u(x,t)\approx \sum_{n=0}^{k}u_{n}(x,t),\;\; \mbox{where} \; \lim_{k\to\infty}\sum_{n=0}^{k}u_{n}(x,t)=u(x,t).\label{eq:y12}
\end{equation}
This method has been successfully applied to a large class of both linear and nonlinear problems \cite{Duan2}. The Adomian decomposition method requires far less work in comparison with  traditional methods. This method considerably decreases the volume of calculations. The decomposition procedure of Adomian easily obtains the solution without linearising the problem by implementing the decomposition method rather than the standard methods. In this approach, the solution is found in the form of a convergent series with easily computed components;
in many cases, the convergence of this series is extremely fast and consequently only a few terms are needed in order to have an idea of how the solutions behave. Convergence conditions of this series have been investigated by several authors, e.g., \cite{Y3,Y4,Y1,Y2}.

\section*{The Mathematical Model of the Burgess Equation}
\label{sn:MM}

\noindent Mathematical modelling of the spread of aggressive brain cancers such as glioblastoma multiforme has been discussed by several authors \cite{Burgess}, \cite{Wein}, \cite{Murray}. It is noteworthy to say that some authors like  \cite{Wein} have included a killing term. In any case, they  describe tumour-growth by using spatiotemporal models that can be read as
\begin{center}
	Rate of change of tumour cell density\\
	= (Diffusion of tumour cells)\\
	+ (Growth of tumour cells)\\-(Killing rate of the same cells)
\end{center}
in mathematical terms,
\begin{eqnarray}
	\frac{\partial \eta(r,t)}{\partial t}=D\frac{1}{r^{2}}\frac{\partial}{\partial r}(r^{2}\frac{\partial \eta(r,t)}{\partial r})+p(t)\eta(r,t)-k(t)\eta(r,t).\label{eq:def3}
\end{eqnarray}
\noindent In this equation,  $\eta(r, t)$ is the concentration of tumour cells at location $r$ at time $t$, $D$ is the diffusion coefficient, {\it i.e.} a measure of the areal speed of the invading glioblastoma cells, $p$ is the rate of reproduction of the glioblastoma cells, and $k$ the killing rate of the same cells. The last term has been used by some authors to investigate the effects of chemotherapy  \cite{Trac}, \cite{Wood}. In this model, the tumour is assumed to have spherical symmetry and the medium through which it is expanding, to be isotropic and uniform.
\noindent We can assume that at the beginning of time (diagnostic time $t_{0}$), the density of cancer cells is $N_0$, {\it i. e.}, $\eta(r_{0},t_{0})=N_0$ and so the equation (\ref{eq:def3}) is
\begin{equation}
	\left\{
	\begin{array}{ll}
		\frac{\partial \eta(r,t)}{\partial t}=D(r,t) \big (\frac{\partial^{2} \eta(r,t)}{\partial r^{2}}+\frac{2}{r}\frac{\partial \eta(r,t)}{\partial r} \big )+[p(t)-k(t)]\eta(r,t),\\
		\eta(r_{0},t_{0})=N_0.
	\end{array}
	\right.\label{eq:y13}
\end{equation}
\noindent The solution of  (\ref{eq:y13}) is given, without many details, in \cite{Burgess} and in  \cite{Murray-0}. They solve this equation for the -non-very realistic- case in which $k(t)\equiv 0$ and  $p(t)$ and $D(r,t)$ are constants. The solution for this case is given by
\begin{equation}
	\eta(r,t)=\frac{N_{0}e^{\{pt-\frac{r^{2}}{4Dt}\}}}{8(\pi Dt)^{\frac{3}{2}}}\label{eq:y14}.
\end{equation}
Using this equation (\ref{eq:y14}), the mentioned authors calculate the expected survival time (in months) for a person that has a  brain tumour modelled using equation  (\ref{eq:y13}).

\begin{figure} [h]
	\begin {center}
	\includegraphics[width=0.8\textwidth]{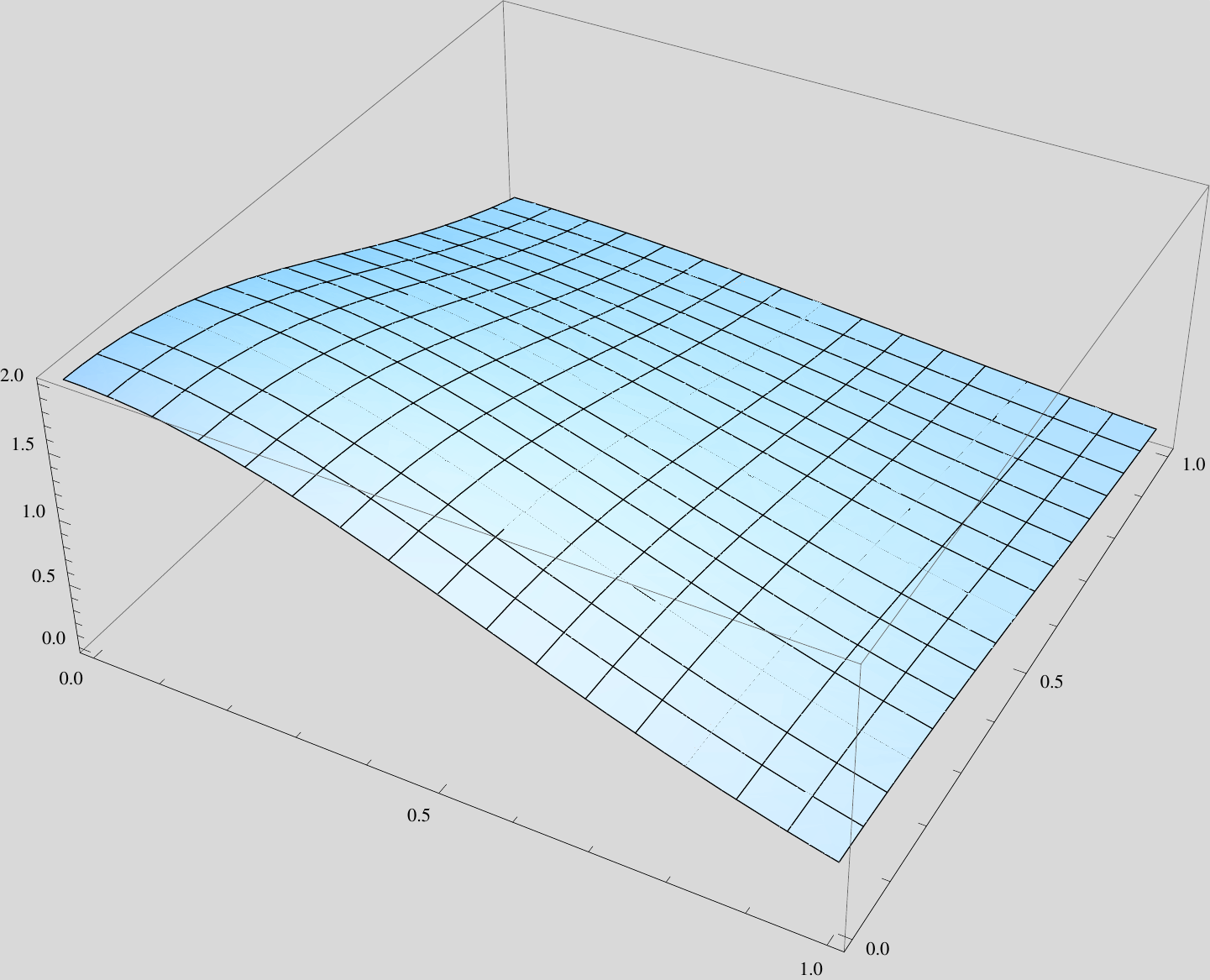}
	\caption{\small  $\omega(r,\tau)=e^{-u}+\frac{1}{2}e^{-2u}$ with $(r,\tau)\in (0,1]\times [0,1]$\label{F2}  }
	\end {center}
\end{figure}

\noindent  Following \cite{Leach}, we  propose the change of variables   $\tau=2Dt$, $u(r,\tau)=r\eta(r,t)$ and  $\omega(r,\tau)=\frac{p-k}{2D}$. Using this change of variables  and keeping  $D$ constant, equation (\ref{eq:y13}) is given by
\begin{equation}
	\frac{\partial u(r,\tau)}{\partial \tau}=\frac{1}{2}\frac{\partial^{2} u(r,\tau)}{\partial r^{2}}+\frac{\omega (r,\tau)}{2D}.\label{eq:y15}
\end{equation}
\noindent In \cite{Roger}, a medical study is presented that stresses the advantages of using combined therapies such as chemotherapy and radiotherapy in the treatment of brain cancer.  Concretely, they present the results of using temozolomide in combination with radiotherapy. The results show a lengthening in the patient life  as a consequence of the tumour size decrease. Mathematically this is traduced as a negative growth of the tumour, in other words, the term $p(r,t)-k(r,t)$ (growth of the cancer cells minus eliminations of cancer cells) is negative.
In present work, we will study the case presented by Roger Stupp {\it et. al.} in \cite{Roger}. Our model will make use of  equation (\ref{eq:y15})  taking $D$ constant and $(r,\tau)\in (0,1]\times [0,1]$ that gives a renormalised time and space intervals. In order to take into account the combined effects of radiotherapy and chemotherapy, we will introduce a term that models the decay (negative growth) of the glioma, $\omega(r,\tau)=e^{-u}+\frac{1}{2}e^{-2u}$, with   $u=u(r,\tau)$.  The decrease of the tumour depending on the position $r$ and  time $\tau$ as it shown in figure \ref{F2}.

\noindent The ADM has been used by several authors to solve linear and non-linear diffusion equations  as well as fractional diffusion equations, some important references can be found in \cite{X8,X9,X6,X1,X2,X3,X4,X5,X7}. In the present work we are interested in the solution of the diffusion equation (\ref{eq:y15})  in which a non-linear source $\omega(r,\tau)$ is modelling the effects of the combined use of radiotherapy and chemotherapy treatment with  {\it Temozolomide} as is reported in \cite{Roger}.

\section*{Solution of a nonlinear model}
\begin{figure} [h]
	\begin {center}
	\includegraphics[width=0.8\textwidth]{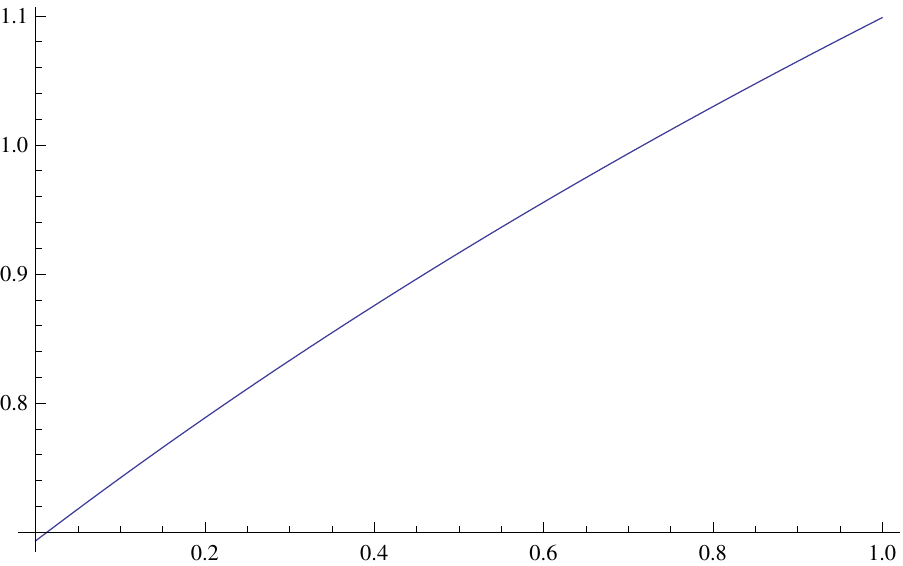}
	\caption{\small Initial growth-profile $u(r,0)=\ln(r+2)$\label{F3}  }
	\end {center}
\end{figure}

\noindent Considering the equation  (\ref{eq:y15}), with $\omega(r,\tau)=e^{-u}+\frac{1}{2}e^{-2u}$ and $D=\frac{1}{2}$ our model will be given by the following non-linear partial differential equation 

\begin{equation}
	\left\{
	\begin{array}{ll}
		\frac{\partial u}{\partial \tau}=\frac{1}{2}\frac{\partial^{2} u}{\partial r^{2}}+e^{-u}+\frac{1}{2}e^{-2u},\\
		u(r,0)=\ln(r+2).
	\end{array}
	\right.\label{eq:y16}
\end{equation}
In equation  (\ref{eq:y16}) we have made the  {\it a priori} assumption that  the initial condition is  $u(r,0)=\ln(r+2)$. This assumption considers  that the initial tumour growth profile is given by $u(r,0)$  in the time we start the annihilation or attenuation of the gliomas by means of some treatment (as chemotherapy).   The initial growth profile is illustrated in figure \ref{F3}.

\noindent Using
$$
A_n(u_{0},u_{1},\ldots,u_{n})= \frac{1}{n!}\frac{d^{n}}{d\alpha^{n}}[N(\sum_{k=0}^{n}\alpha^{k}u_{k})]\rvert_{\alpha=0}\quad n\geq 0$$
to calculate the Adomian polynomials, we have :
$$A_{0}(u_{0})=N(u_{0})=e^{-u_{0}}+\frac{1}{2}e^{-2u_{0}}$$
$$A_{1}(u_{0},u_1)=N'(u_{0})u_{1}=-u_{1}e^{-u_{0}}-u_{1}e^{-2u_{0}}$$
$$\begin{array}{lcl} A_{2}(u_{0},u_1, u_2) & = & N''(u_{0})\frac{u_{1}^{2}}{2!}+N'(u_0)u_{2}\\ & =& \frac{u_{1}^{2}}{2}(e^{-u_{0}}+2e^{-2u_{0}})+u_{2}(-e^{-u_{0}}-e^{-2u_{0}})\end{array}$$
$$A_{3}(u_{0},u_1, u_2, u_3)  =  N'''(u_{0})\frac{u_{1}^{3}}{3!}+N''(u_0)u_{1}u_{2}+N'(u_0)u_{3}$$ $$ =\frac{u_{1}^{3}}{6}(-e^{-u_{0}}-4e^{-2u_{0}})+u_{1}u_{2}(e^{-u_{0}}+2e^{-2u_{0}})-u_{3}(e^{-u_{0}}+e^{-2u_{0}})$$
$$\vdots . $$
\noindent Using the  sequence for  $\{A_n\}_{n=0}^{\infty}$ and the recurrence relation given in (\ref{eq:y11})  we can calculate  $\{u_n\}$, in this way:
$$u_{0}=\ln(r+2) $$
$$u_{1}=\int_{0}^{t}[-\frac{1}{2(r+2)^{2}}+e^{-u_{0}}+\frac{1}{2} e^{-2u_{0}}]d\hat{t}=\frac{t}{r+2}$$
$$\vdots .$$

\noindent The partial sums of the  Adomian series are  
$$S_{0}=u_{0}=\ln(r+2) $$
$$S_{1}=u_{0}+u_{1}=\ln(r+2)+\frac{\tau}{r+2} $$
$$S_{2}=u_{0}+u_{1}+u_{2}=\ln(r+2)+\frac{\tau}{r+2}-\frac{\tau^{2}}{2(r+2)^{2}} $$
$$S_{3}=u_{0}+u_{1}+u_{2}+u_{3}=\ln(r+2)+\frac{\tau}{r+2}-\frac{\tau^{2}}{2(r+2)^{2}}+ \frac{\tau^{3}}{3(r+2)^{3}}$$
$$\vdots $$
$$S_{m}=u_{0}+u_{1}+\ldots+u_{m}=\ln(r+2)+\frac{\tau}{r+2}-\frac{\tau^{2}}{2(r+2)^{2}}+\ldots+\frac{(-1)^{m+1}\tau^{m}}{m(r+2)^{m}}$$
and taking into account the equation  (\ref{eq:y12}), we have

\begin{equation}
	u(r,\tau)=\ln(r+2)+\frac{\tau}{r+2}-\frac{\tau^{2}}{2(r+2)^{2}}+\ldots+\frac{(-1)^{m+1}\tau^{m}}{m(r+2)^{m}}+\cdots\label{eq:y18}
\end{equation}
taking the sum of the first  terms, we can see that the  above series converges to $\ln(\frac{\tau+r+2}{r+2})$. Then, using   (\ref{eq:y18}) we have
\begin{equation}
	u(r,\tau)=\ln(r+2)+\ln(\frac{\tau+r+2}{r+2})=\ln(r+\tau+2).\label{eq:y19}
\end{equation}

\begin{figure} [h]
	\begin {center}
	\includegraphics[width=0.8\textwidth]{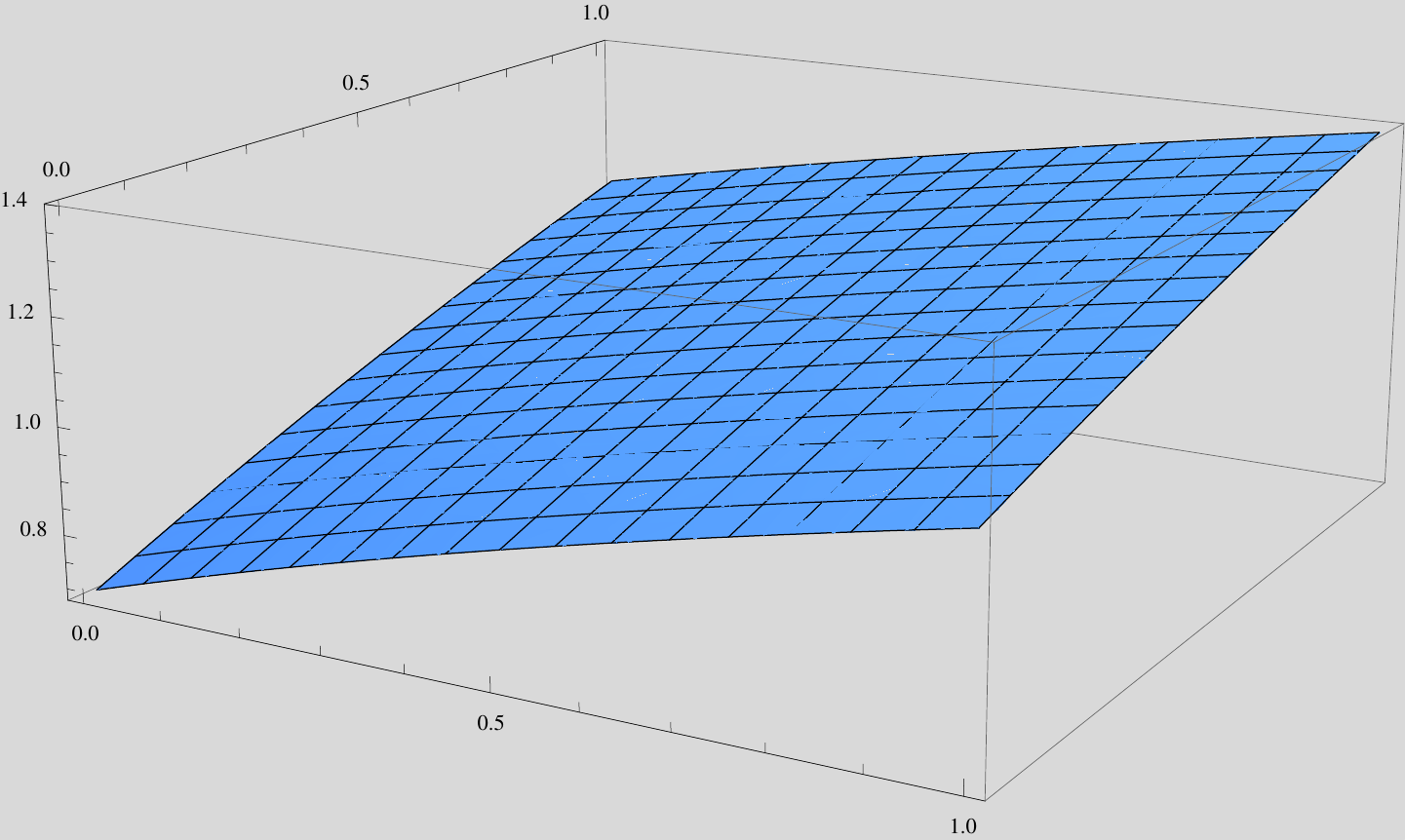}
	\caption{\small $u(r,\tau)=\ln(r+\tau+2)$ with $(r,\tau)\in (0,1]\times [0,1]$\label{F4}  }
	\end {center}
\end{figure}

\noindent It is easy to verify that  $u(r,\tau)$ given by (\ref{eq:y19}) is a solution of the initial value problem  (\ref{eq:y16}). With this solution we can given the density of cancer cells in every point of  $(r,\tau)\in (0,1]\times [0,1]$. In figure  \ref{F4},  we can observe   the approximate-linear  tumour growth-profile after the patient is under chemotherapy treatment with  {\it Temozolomide} in contrast  with the fast exponential growth given by  (\ref{eq:y14}) and corresponding to free-growth tumour.  The free-growth tumour profile  was shown in \cite{Burgess}, in which  the value of  $\eta(r,t)$ is given for different values of  the parameters $D$,  $p$ and $k(t)=0$. \\

\noindent In order to see the effect of medical treatment, we can compare the  radius of the tumour under medical treatment versus the radius of the untreated  tumour. Using the solution (\ref{eq:y14}) of the Burgess linear partial equation and solving for $r$  
(also see \cite{Murray-0})  that accounts for free growth of an untreated tumour , we obtain
\begin{equation}\label{eq:y-20}
	r_{lin}=2t\sqrt{Dp}\sqrt{\left| 1-\frac{1}{pt}\ln[\frac{C_0}{N_0}(4\pi Dt)^{\frac{3}{2}}]\right|}.
\end{equation}
Solving for $r$ in  the solution of the non-linear  equation (\ref{eq:y19})  that accounts for case in which we have proliferation and annihilation of tumour cells due to medical treatment,  we have 
\begin{equation}\label{eq:y-21}
	r_{nonlin}=e^{N_{0}}-\tau-2
\end{equation}
If we take $D=0.5$, $p=1.5$ and $\frac{C_0}{N_0}=4000$ (see \cite{Murray-0}) and recalling that $\tau=2Dt$, then  $t=\tau$.
Using this values in equation (\ref{eq:y-20}) and (\ref{eq:y-21})
we obtain the following table 

\begin{figure} [h!]
	\begin {center}
	\includegraphics[width=12cm,height=6cm]{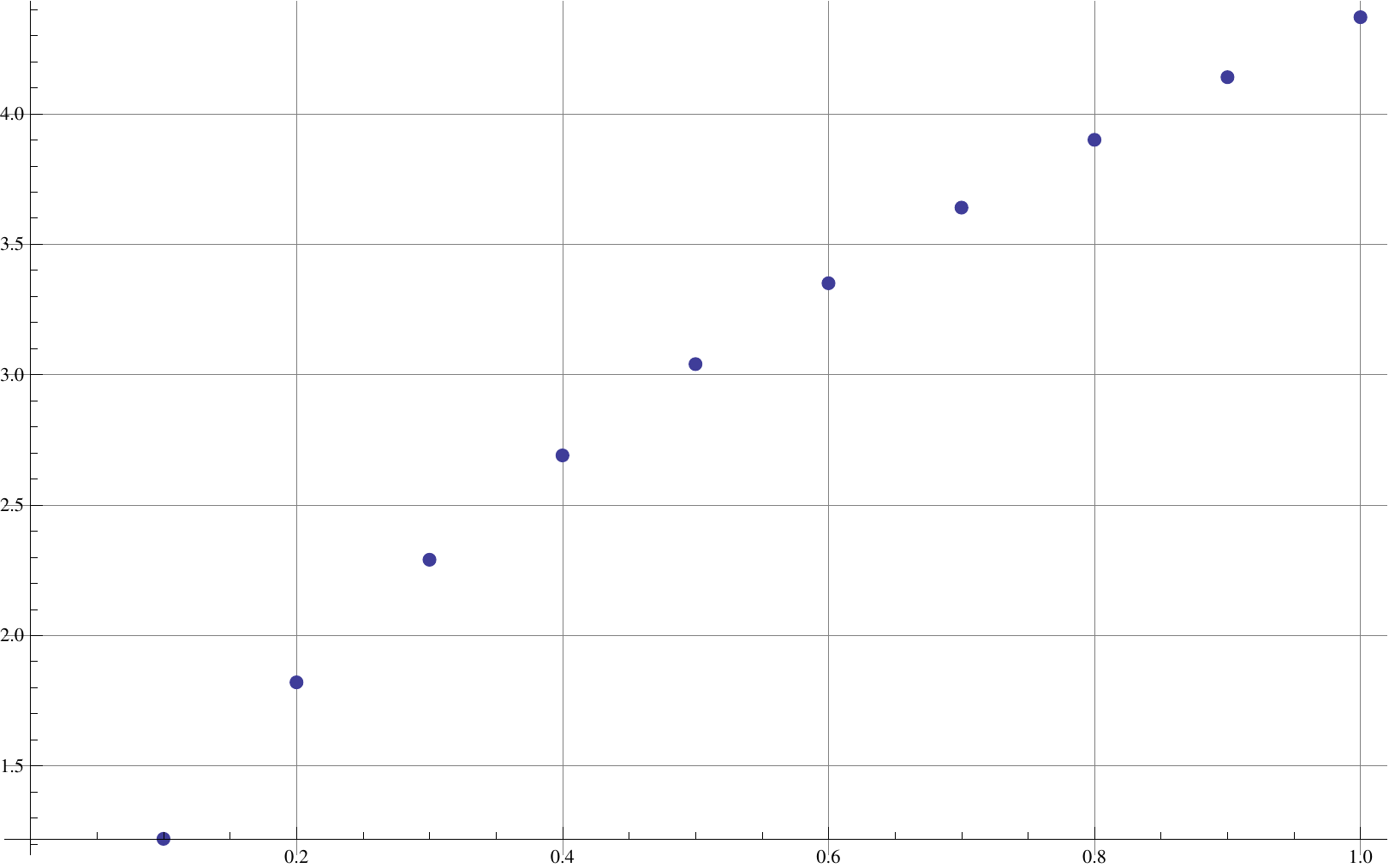}
	\caption{\small  Tumour's radius  $r_{lin}$ versus time $t\in (0,1]$ for equation (20)  \label{F5}  }
	\end {center}
\end{figure}

\begin{figure} [h!]
	\begin {center}
	\includegraphics[width=12cm,height=6cm]{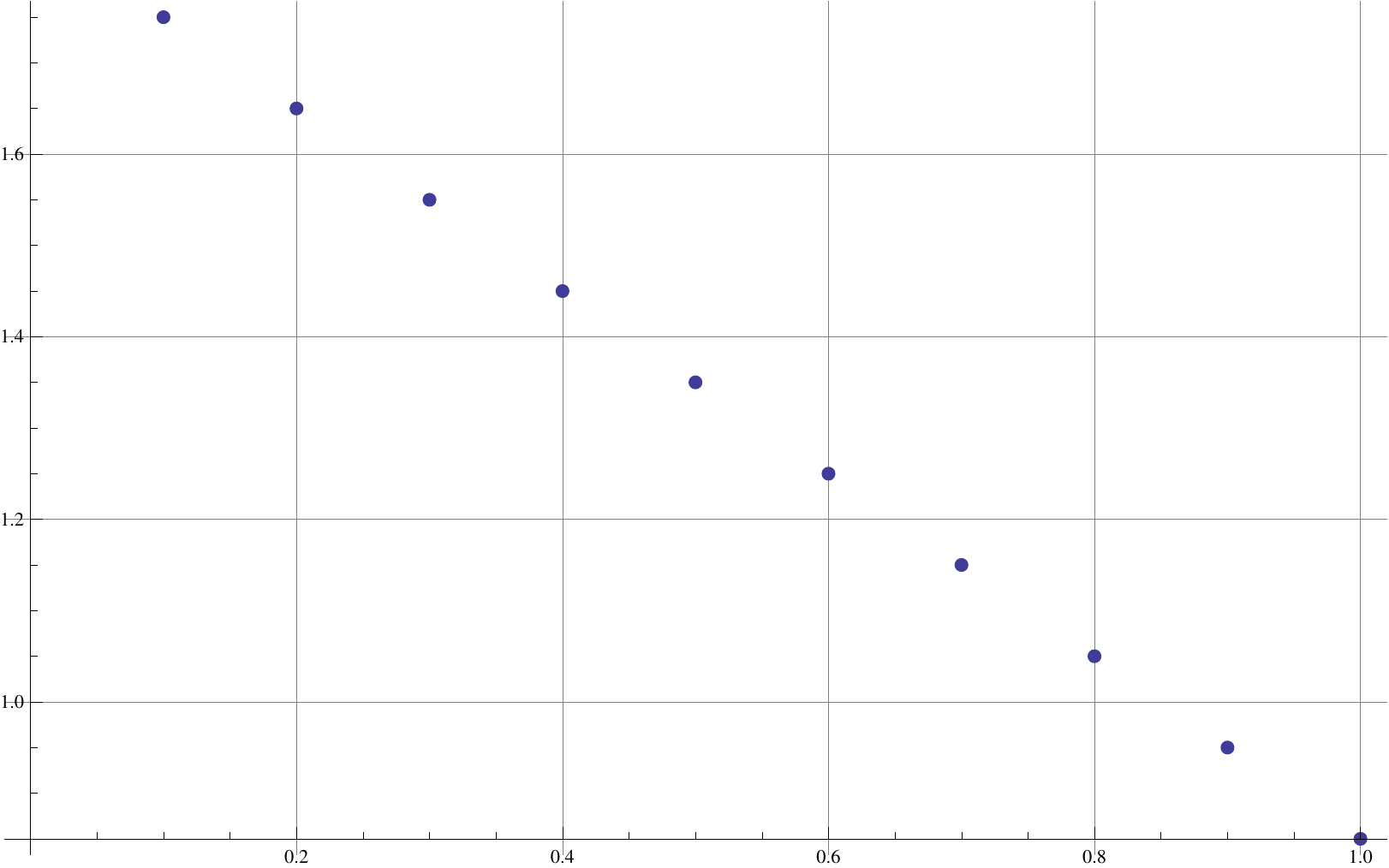}
	\caption{\small  Tumour's radius  $r_{nonlin}$ versus time $t\in (0,1]$ for equation (21) \label{F6} }
	\end {center}
\end{figure}

\begin{table}[h]
	\begin{center}
		\begin{tabular}{ |l|l|l| }
			\hline
			$t(years)$ & $r_{lin}(cm)$ & $r_{nonlin}(cm)$  \\ \hline
			0.1 & 1.22 & 1.75 \\ \hline
			0.2 & 1.82 & 1.65 \\ \hline
			0.3 & 2.29 & 1.55\\ \hline
			0.4 & 2.69 & 1.45 \\ \hline
			0.5 & 3.04 & 1.35 \\ \hline
			0.6 & 3.35 & 1.25\\ \hline
			0.7 & 3.64 & 1.15 \\ \hline
			0.8 & 3.90 & 1.05 \\ \hline
			0.9 & 4.14 & 0.95\\ \hline
			1.0 & 4.37 & 0.85\\ \hline
		\end{tabular}
	\end{center}
	\caption{Radius growth in untreated tumour versus radius growth in treated tumour.}
	\label{t1}   
\end{table}
\noindent Using the data of table  \ref{t1} we obtain figures \ref{F5} and \ref{F6}. From the figure \ref{F5}, we observe that the radius of the untreated tumour grows as predicted in (\ref{eq:y14})  meanwhile we observe,  in figure \ref{F6}, that the treated tumour's radius decreases with time.

\section*{Summary}
\noindent  In this work we have proposed a model for  cerebral tumour  (glioblastoma) growth under medical treatment. Taking the Burgess equation as departing point, we considered  additional  non-linear terms that represent the dynamics of proliferation and annihilation of gliomas resulting from medical care as suggested in the clinical study done by R. Stupp \cite{Roger}. The effect of the medical treatment on the tumour is represented by a non-linear term. The final model that describes the proliferation  and annihilation of tumour cells is represented by a partial non-linear differential equation that is solved using the Adomian Decomposition Method (ADM). Finally, as is observed in  table  \ref{t1} and figures  \ref{F5} and \ref{F6} our model and the solution given using the ADM appropriately  models the effects of the combined therapies. By means of a proper use of parameters, this model could be used for calculating doses in radiotherapy and chemotherapy

\section*{Acknowledgments}
We would like to thank anonymous referees for their constructive comments and suggestions that helped to improve the paper.

\section*{Appendix: Adomian polynomials}
In this appendix we will  deduce equation  (\ref{eq:y9}) that accounts for every term in the succession of the  Adomian Polynomials assuming  the following hypotheses stated in \cite{Y2}:

\begin{itemize}
	
	\item[(i)] the series solution, $u=\sum_{n=0}^{\infty}u_n$ ,  of the problem given in equation  (\ref{eq:y1}) is absolutely convergent,
	\item[(ii)] The non-linear function $N(u)$ can be expressed by means of a power series whose radio of convergence is infinite, that is
	\begin{equation}\label{2.1}
		N(u)=\sum_{n=0}^{\infty}N^{(n)}(0)\frac{u^n}{n!}\ ,\ \ \ \ |u|<\infty.
	\end{equation}
	
\end{itemize}

\noindent Assuming  the above hypotheses, the series whose terms are the Adomian Polynomials  $\{A_{n}\}_{n=0}^{\infty}$ results to be a generalisation of the Taylor's series
\begin{equation}\label{2.2}
	N(u)=\sum_{n=0}^{\infty}A_n(u_0,u_1,\ldots,u_n)=\sum_{n=0}^{\infty}N^{(n)}(u_0)\frac{(u-u_0)^n}{n!}.
\end{equation}

\noindent Is worthy to note that (\ref{2.2})  is a rearranged expression  of the series (\ref{2.1}), and note that, due to hypothesis,  this series is convergent. 
Consider now, the parametrisation proposed by G. Adomian in  \cite{ADM} given by
\begin{equation}\label{2.3}
	u_\lambda(x,t)=‎‎\sum_{n=0}^{\infty‎}u_n(x,t)f^n(\lambda),
\end{equation}

\noindent where $\lambda$ is a parameter in  $\mathbb{R}$ and  $f$ is a complex-valued function such that  $|f|<1$. With this choosing of  $f$ and using the hypotheses above stated, the series (\ref{2.3}) is absolutely convergent.\\
\noindent Substituting  (\ref{2.3}) in (\ref{2.2}), we obtain 
\begin{equation}\label{2.5}
	N(u_\lambda)=\sum_{n=0}^{\infty}N^{(n)}(u_0)\frac{\left(\sum_{j=1}^{\infty‎}u_j(x,t)f^j(\lambda)\right)^n}{n!}\ .
\end{equation}
Due to the absolute convergence of 
\begin{equation}\label{et1}
	\sum_{j=1}^{\infty‎}u_j(x,t)f^j(\lambda),
\end{equation}  
we can rearrange  $N(u_\lambda)$ in order to obtain the series of the form $\sum_{n=0}^{\infty}A_nf^n(\lambda)$. Using  (\ref{2.5}) we can obtain the coefficients  $A_k$ de $f^k(\lambda)$, and finally we  deduce  the Adomian's polynomials. That is,
\begin{eqnarray}\label{2.6}
	N(u_\lambda)&=&N(u_0)+N^{(1)}(u_0)\left(u_1f(\lambda)+u_2f^2(\lambda)+u_3f^3(\lambda)+\ldots\right)\nonumber\\&&+\frac{N^{(2)}(u_0)}{2!}\left(u_1f(\lambda)+u_2f^2(\lambda)+u_3f^3(\lambda)+\ldots\right)^2\nonumber\\&&+\frac{N^{(3)}(u_0)}{3!}\left(u_1f(\lambda)+u_2f^2(\lambda)+u_3f^3(\lambda)+\ldots\right)^3+\ldots\nonumber\\&&
	+\frac{N^{(4)}(u_0)}{4!}\left(u_1f(\lambda)+u_2f^2(\lambda)+u_3f^3(\lambda)+\ldots\right)^3+\ldots\nonumber\\ 
	&=&N(u_0)+N^{(1)}(u_0)u_1f(\lambda)+\left(N^{(1)}(u_0)u_2+N^{(2)}(u_0)\frac{u_1^2}{2!}\right)f^2(\lambda)\nonumber\\&&
	+\left(N^{(1)}(u_0)u_3+N^{(2)}(u_0)u_1u_2+N^{(3)}(u_0)\frac{u_1^3}{3!}\right)f^3(\lambda)+\ldots \nonumber\\
	&=&\sum_{n=0}^{\infty}A_n(u_0,u_1,\ldots,u_n)f^n(\lambda).
\end{eqnarray}
Using equation  (\ref{2.6}) making  $f(\lambda)=\lambda$ and taking derivative at both sides of the equation, we can make the following identification\\
\noindent $A_0(u_0)=N(u_0)$\\
$A_1(u_0,u_1)=N'(u_0)u_1$\\
$A_2(u_0,u_1,u_2)=N'(u_0)u_2+\frac{u_1^2}{2!}N''(u_0)$\\
$A_3(u_0,u_1,u_2,u_3)=N'(u_0)u_3+N''(u_0)u_1u_2+\frac{u_1^3}{3!}N'''(u_0)$\\
$A_{3}(u_0,\ldots,u_4)=u_4N'(u_0)+(\frac{1}{2!}u_{2}^{2}+u_{1}u_{3})N''(u_0)+\frac{u_{1}^{2}u_{2}}{2!}N'''(u_0)+\frac{u_{1}^{4}}{4!}N^{(iv)}(u_0)$\\
$\vdots$\\
Hence we have obtain equation  (\ref{eq:y9}):
\begin{equation}
	A_n(u_0,u_1,\ldots, u_n)= \frac{1}{n!}\frac{d^{n}}{d\lambda^{n}}[N(\sum_{k=0}^{n}\lambda^{k}u_{k})]\rvert_{\lambda=0}.\label{2.7}
\end{equation}



\begin{thebibliography}{}
	%
	%
	
	\bibitem{Y3}
	Abbaoui, K., Cherruault, Y.: Convergence of Adomian's method applied to differential equations. Comput. Math. Appl. {\bf 28}(5), 103-109 (1994)
	
	\bibitem{Y4}
	Abbaoui, K., Cherruault, Y.: New ideas for proving convergence of decomposition methods.  Comput. Math. Appl. {\bf 29}(7), 103-108 (1995)
	
	\bibitem{ADM}
	Adomian, G.: Nonlinear stochastic operator equations. Orlando: Academic Press, (1986)
	
	\bibitem{ADM2}
	Adomian, G.: A review of the decomposition method in applied mathematics. J. Math. Anal. Appl. {\bf 135}(2), 501-544 (1988)
	
	\bibitem{ADM1}
	Adomian, G., Rach, R.: Nonlinear stochastic differential delay equations.  J. Math. Anal. Appl. {\bf 91}, 94-101 (1983)
	
	
	\bibitem{Leach} 
	Andriopoulos, K.,  Leach, P. G. L.: A common theme in applied mathematics: an equation connecting applications in economics, medicine and physics. South African Journal of Sciences. {\bf 102}, 66-72 (2006)
	
	\bibitem{Burgess}
	Burgess, P. K., {\it et. al.} : The interaction of growth rates and diffusion
	coefficients in a three-dimensional mathematical model of gliomas.
	J. of Neuropathology and Experimental Neurology. {\bf 56}, 704-713 (1997)
	
	\bibitem{Y1}
	Cherruault, Y.: Convergence of Adomian's method.  Kybernetes. {\bf 18}(2), 31-38 (1989)
	
	\bibitem{Y2}
	Cherruault, Y.,  Adomian, G.: Decomposition methods: a new proof of convergence. Math. Comput. Modelling. {\bf 18}(12), 103-106 (1993)
	
	
	\bibitem{X8}
	Das, S.: Generalized dynamic systems solution by decomposed physical reactions. Int.  J. of Applied Math. and Statistics. {\bf 17}, 44-75 (2010)
	
	
	\bibitem{X9}
	Das, S.:  Solution of extraordinary differential equation with physical reasoning by obtaining modal reaction. Series Modelling and Simulation in Engineering. {\bf 2010}, 1-19  (2010). DOI:10.1155/2010/739675
	
	\bibitem{X6}
	Das, S.: Functional Fractional Calculus, 2nd Edition, Springer-Verlag  (2011)
	
	\bibitem{Duan2}
	Duan, J.S., Rach, R., Wazwaz, A. M.: A new modified Adomian decomposition method for higher-order nonlinear dynamical systems. Comput.  Model.  Eng.  Sci. (CMES) {\bf 94}(1), 77-118 (2013)
	
	\bibitem{May}
	Mayfield Clinic Homepage
	http://www.mayfieldclinic.com/
	
	
	\bibitem{Murray-0}  
	Murray, J. D.: Glioblastoma brain tumors: estimating the time from brain tumor initiation and resolution of a patient survival anomaly after similar treatment protocols.  J. of Biological Dynamics. {\bf 6}, suppl. 2, 118-127 (2012). DOI: 10.1080/17513758.2012.678392
	
	
	\bibitem{Murray} 
	Murray,  J. D.: Mathematical Biology II: Spatial Models and Biomedical Applications. 3rd Ed. Springer-Verlag, New York  (2003)
	
	\bibitem{X1}
	Saha Ray,  S.,  Bera, R.  K.:  Analytical solution of Bagley Torvik equation by Adomian's decomposition method. Appl. Math. Comp. {\bf 168}, 389-410 (2005)
	
	\bibitem{X2}
	Saha Ray, S., Bera, R.  K.:  An approximate solution of nonlinear fractional differential equation by Adomian's decomposition method.  Applied Math. Computation. {\bf 167}, 561-71 (2005)
	
	\bibitem{X3}
	Saha Ray, S.:  Exact solution for time fractional diffusion wave equation by decomposition method.  Physics Scripta. {\bf 75},  53-61 (2007)
	
	\bibitem{X4}
	Saha Ray,  S.:  A new approach for the application of Adomian decomposition method for solution of fractional space diffusion equation with insulated ends.  Applied Math.  and Computations. {\bf 202}(2), 544-549 (2008)
	
	
	\bibitem{X5}
	Saha Ray, S., Bera, R.  K.:  Analytical solution of dynamic system containing fractional derivative of order one-half by Adomian decomposition method. ASME J. of Applied Mechanic. {\bf 72}(1)  (2005)
	
	
	\bibitem{X7}
	Sardar, T., Saha Ray, S., Bera, R.  K., Biswas B. B.,  Das, S.: The solution of coupled fractional neutron diffusion equations with delayed neutron. Int. J. of Nuclear Energy Science and Tech. {\bf 5}(2), 105-113  (2010)
	
	
	\bibitem{Sch}
	Schiffer, D. Annovazzi,  L., Caldera,  V., Mellai, M.: On the origin and growth of gliomas.  Anticancer Research. {\bf 30}(6),  1977–1998, (2010)
	
	
	
	\bibitem{Roger}
	Stupp, R.,  {\it et. al.}: Radiotherapy plus concomitant and adjuvant temozolomide for glioblastoma. New Engl. Journal of Med. {\bf 352}, 987-996 (2005)
	
	
	\bibitem{Trac}
	Tracqui, P.,  {\it et.  al.}: A mathematical model of glioma growth: the effect of chemotherapy on spatio-temporal growth. Cell. Proliferation.  {\bf 28}, 17-31 (1995)
	
	\bibitem{Wein}
	Wein, L., Koplow, D.:  Mathematical modeling of brain cancer to identify promising combination treatments.  Preprint, D Sloan School of Management, MIT (1999)
	
	\bibitem{Wen}
	Wenhai, C., Zhengyi L.: An algorithm for Adomian decomposition method. Applied Mathematics and Computation {\bf 159}, 221-235 (2004)
	
	\bibitem{Wood}
	Woodward, D. E.,  {\it et. al.}:  A mathematical model of glioma growth: the effect of extent of surgical resection; Cell. Proliferation, {\bf 29}, 269-288 (1996)
	
	
	
	
	
\end{thebibliography}
\end{document}